\documentclass[superscriptaddress,prl,12pt]{revtex4}

\usepackage{natbib}

\usepackage{anysize}
\marginsize{2.3cm}{2.3cm}{1.5cm}{1.5cm}
\usepackage{graphicx}
\usepackage{amsmath}
\usepackage[bookmarks = false]{hyperref}

\begin{document}

\title{Quantum teleportation between remote atomic-ensemble quantum memories}

\author{Xiao-Hui Bao}
\affiliation{Hefei National Laboratory for Physical Sciences at
Microscale and Department of Modern Physics, University of Science
and Technology of China, Hefei, Anhui 230026, China}
\affiliation{Physikalisches Institut der Universitaet Heidelberg,
Philosophenweg 12, Heidelberg 69120, Germany}

\author{Xiao-Fan Xu}
\affiliation{Physikalisches Institut der Universitaet Heidelberg,
Philosophenweg 12, Heidelberg 69120, Germany}

\author{Che-Ming Li\footnote{Present address: Department of Engineering Science and Supercomputing Research Center, National Cheng Kung University, Tainan 701, Taiwan}}
\affiliation{Physikalisches Institut der Universitaet Heidelberg,
Philosophenweg 12, Heidelberg 69120, Germany}

\author{Zhen-Sheng Yuan}
\affiliation{Hefei National Laboratory for Physical Sciences at
Microscale and Department of Modern Physics, University of Science
and Technology of China, Hefei, Anhui 230026, China}
\affiliation{Physikalisches Institut der Universitaet Heidelberg,
Philosophenweg 12, Heidelberg 69120, Germany}

\author{Chao-Yang Lu$^\dag$}
\affiliation{Hefei National Laboratory for Physical Sciences at
Microscale and Department of Modern Physics, University of Science
and Technology of China, Hefei, Anhui 230026, China}

\author{Jian-Wei Pan\footnote{Correspondence and requests for materials may be addressed to C.Y.L. (cylu@ustc.edu.cn) or J.-W.P (pan@ustc.edu.cn).}}
\affiliation{Hefei National Laboratory for Physical Sciences at
Microscale and Department of Modern Physics, University of Science
and Technology of China, Hefei, Anhui 230026, China}
\affiliation{Physikalisches Institut der Universitaet Heidelberg,
Philosophenweg 12, Heidelberg 69120, Germany}

\maketitle

\textbf{
Quantum teleportation and quantum memory are two crucial elements for large-scale quantum networks. With the help of prior distributed entanglement as a ``quantum channel'', quantum teleportation provides an intriguing means to faithfully transfer quantum states among distant locations without actual transmission of the physical carriers [C.H. Bennett \textit{et al.} (1993) Phys. Rev. Lett. 70, 1895-1899]. Quantum memory enables controlled storage and retrieval of fast-flying photonic quantum bits with stationary matter systems, which is essential to achieve the scalability required for large-scale quantum networks. Combining these two capabilities, here we realize quantum  teleportation between two remote atomic-ensemble quantum memory nodes, each composed of $\mathbf{\sim}\,$$\mathbf{10^8}$ rubidium atoms and connected by a 150-meter optical fiber. The spinwave state of one atomic ensemble is mapped to a propagating photon, and subjected to Bell-state measurements with another single photon that is entangled with the spinwave state of the other ensemble. Two-photon detection events herald the success of teleportation with an average fidelity of $\mathbf{88(7)\%}$. Besides its fundamental interest as the first teleportation between two remote macroscopic objects, our technique may be useful for quantum information transfer between different nodes in quantum networks and distributed quantum computing.}

\vspace{0.5cm}

Single photons are so far the best messengers for quantum networks as they are naturally propagating quantum bits (qubits) and have very weak coupling to the environment \cite{Gisin2002, physicalreport}. However, due to the inevitable photon loss in the transmission channel, the quantum communication is limited currently to a distance of about 200 kilometers \cite{Stucki2009, Liu2010}. To achieve scalable long-distance quantum communication\cite{Briegel1998, Duan2001}, quantum memories are required \cite{Kimble2008, luming_rmp, gisin_rmp, simon2010}, which coherently convert a qubit between light and matter efficiently on desired time points so that operations can be appropriately timed and synchronized. The connection of distant matter qubit nodes and transfer of quantum information between the nodes can be done by distributing atom-photon entanglement through optical channels and quantum teleportation \cite{teleportation_Bennett}.

Optically-thick atomic ensemble has been proved to be an excellent candidate for quantum memory \cite{kuzmich, lukin, Choi2008, Chou2007, Yuan2008, Radnaev2010}, with promising experimental progress including the entanglement between two atomic ensembles \cite{macroscopic, Chou2005}, generation of nonclassical fields \cite{kuzmich, lukin}, efficient storage and retrieval of photonic qubits \cite{Choi2008}, sub-second storage time \cite{Radnaev2010}, and demonstration of a preliminary quantum repeater node \cite{Chou2007, Yuan2008}. Quantum teleportation has been demonstrated with single photons \cite{Bouwmeester1997, Martini1997,TeleCWFurusawa1998}, from light to matter \cite{Sherson2006, Chen2008}, and between single ions \cite{Riebe2004, Barrett2004, Olmschenk2009}. However, quantum teleportation between remote atomic ensembles has not been realized yet.

\begin{figure*}[htbp]
\centering
\includegraphics[width=0.9\textwidth]{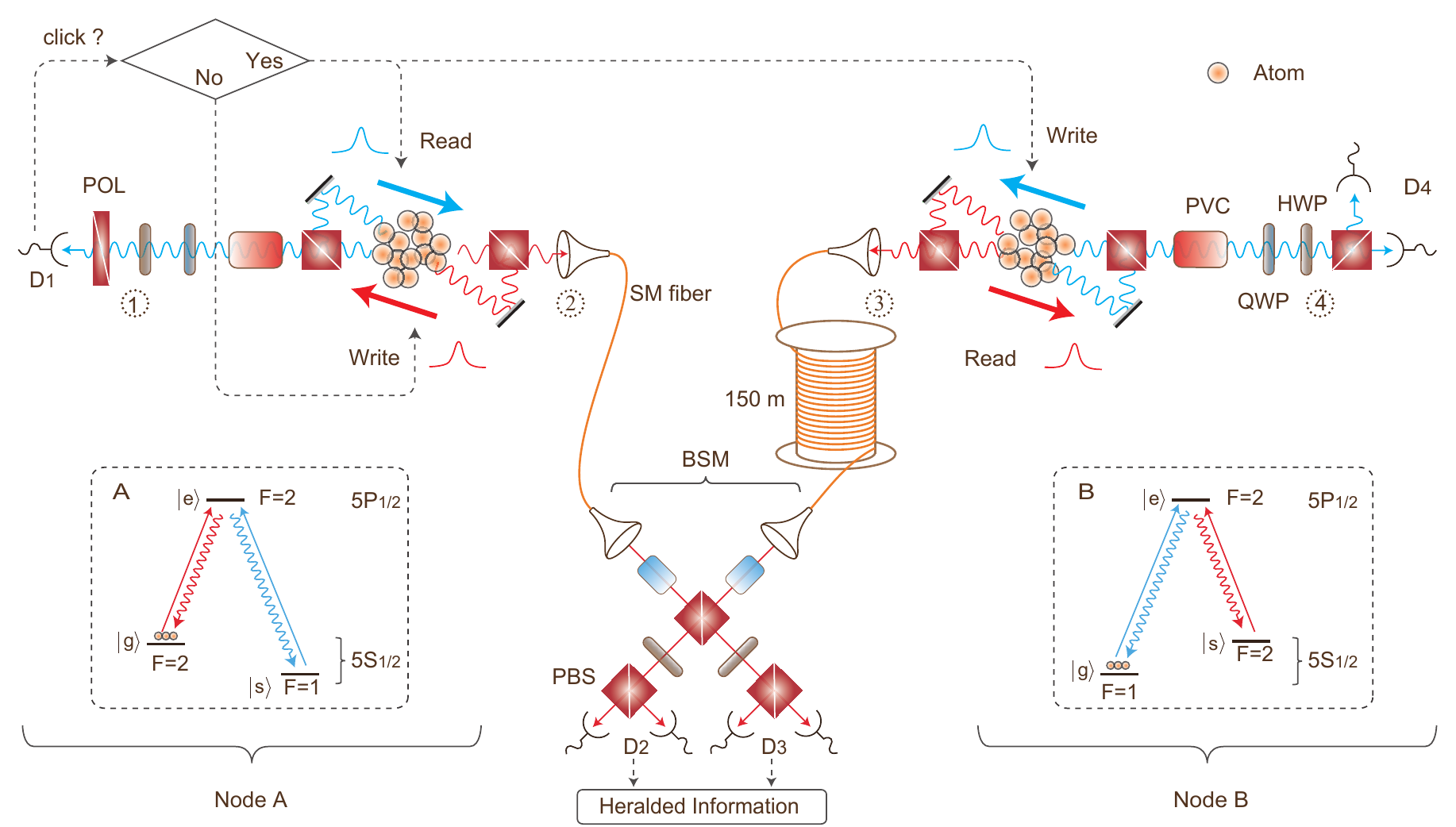}
\caption{\textbf{The experiment setup for quantum teleportation between two remote atomic ensembles}. All the atoms are first prepared at the ground state $|g\rangle$. The spinwave state of atomic ensemble A is prepared through the repeated write process. Within each write trail, with a small probability, entanglement between the spinwave vector and the momentum of the write-out photon is created. A polarizing beam-splitter (PBS) converts the photon's momentum to its polarization. A click in $D_1$ heralds a successful state preparation for ensemble A. Conditioned on a successful preparation, a write pulse is applied on atomic ensemble B creating a pair of photon-spinwave entanglement $|\Phi^+\rangle_{3B}$. The scattered photon 3 travels through a 150 m-long single-mode fibre and subjects to a Bell-state measurement together with the read-out photon 2 from the atomic ensemble A. A coincidence count between detector $D_2$ and $D_3$ heralds the success of teleportation. To verify the teleported state in atomic ensemble B, we convert the spinwave state to the polarization state of photon 4 by applying the read pulse. Photon 4 is measured in arbitrary basis with the utilization of a quarter-waveplate (QWP), a half-waveplate (HWP) and a PBS. The leakage of the write and read pulse into the single-photon channels are filtered out using the pumping vapor cells (PVC). The $\Lambda$-type level schemes used for both ensembles are shown in the insets.
}\label{tele:setup}
\end{figure*}

In this Article we report the first teleportation experiment between two atomic-ensemble quantum memories. The layout of our experiment is shown in Fig. 1. Two atomic ensembles of $^{87}$Rb are created using magneto-optical trap and locate at two separate nodes. The radius of each ensemble is $\sim\,$1 mm. We aim to teleport a single collective atomic excitation (spinwave state) from ensemble A to B which are linked by a 150-meter long optical fibre and physically separated by $\sim\,$$0.6$ meters. The spinwave state can be created through the process of electromagnetically induced transparency \cite{Fleischhauer2005} or weak Raman scattering \cite{Duan2001}, and can be written as
\begin{align}
|\mathrm{dir}\rangle =\frac{1}{\sqrt{N}}\sum_{j}e^{i \mathbf{k}_{\mathrm{dir}}\cdot \mathbf{r}_{j}}|g...s_{j}...g\rangle,
\end{align}
where ``dir" refers to the direction of the spinwave vector $\mathbf{k}_{\mathrm{dir}}$, $\mathbf{r}_{j}$ refers to the coordinate of $j$-th atom, and $N$ refers to the number of atoms. The atoms are in a collective excited state with only one atom excited to the state $|s\rangle$ and delocalized over the whole ensemble. The spinwave can be converted to a single photon with a high efficiency ($>70\%$ has been reported in \cite{Simon2007, Bao2012}) due to the collective enhancement effect \cite{Duan2001, Fleischhauer2005}.

Our experiment starts with initializing the atomic ensemble A in an arbitrary state to be teleported
$|\psi\rangle_{A}=\alpha |\uparrow\,\rangle_A + \beta |\downarrow\,\rangle_A$, where $\uparrow$ (up) and $\downarrow$ (down) refer to the directions of the spinwave vector relative to the write direction in Fig. 1, and $\alpha$ and $\beta$ are arbitrary complex numbers fulfilling $|\alpha|^2 + |\beta|^2 = 1$. To do so, the method of remote state preparation \cite{RSP_Bennett} is used. By applying a write pulse, we first create a pair of entanglement between the spinwave vector and the momentum (emission direction) of the write-out photon (photon 1 in Fig. 1) through Raman scattering \cite{Chen2007}. The momentum degree of the write-out photon is later converted to the polarization degree by a polarizing beam-splitter (PBS). In this way we create the entanglement between the spinwave state of the ensemble and the polarization of the write-out photon. The created atom-photon entangled state can be written as $|\Psi^-\rangle_{1A} = 1/\sqrt{2}(|H\rangle_1|\uparrow\,\rangle_A - |V\rangle_1|\downarrow\,\rangle_A)$. Next, we perform a projective measurement of photon 1 in the basis of $|\psi\rangle_1 \,/\, |\psi^{\perp}\rangle_1$ where $|\psi\rangle_1 = \alpha |H\rangle_1 + \beta |V\rangle_1$ and $|\psi^{\perp}\rangle_1 = \beta^\ast |H\rangle_1 - \alpha^\ast |V\rangle_1$. Due to the anti-correlation nature of $|\Psi^-\rangle$ in an arbitrary basis, if the measurement result gives $|\psi^{\perp}\rangle_1$, we can infer that the state of ensemble A will be projected to $|\psi\rangle_{A}$. Experimentally, we use a combination of a quarter-wave plate, a half-wave plate and a polarizer to measure photon 1 in an arbitrary basis. Due to the probabilistic character in the Raman scattering process, the excitation probability for each write pulse is made to be sufficiently low ($\sim$$\,0.003$) in order to suppress the double-excitation probability. Therefore, the write process needs to be repeated many times in order to prepare the atomic state successfully. The storage lifetime for prepared states is measured to be 129 $\mu$s, which is mainly limited by motion induced dephasing \cite{Zhao2009}.
In our experiment we select the following six initial states to prepare: $|\uparrow\,\rangle_A$, $|\downarrow\,\rangle_A$, $|+\rangle_A$, $|-\rangle_A$, $|R\rangle_A$ and $|L\rangle_A$ with $|\pm\rangle_A = 1/\sqrt{2}(|\uparrow\,\rangle_A \pm |\downarrow\,\rangle_A)$ and $|R/L\rangle_A = 1/\sqrt{2}(|\uparrow\,\rangle_A \pm i|\downarrow\,\rangle_A)$ by projecting photon 1 into the corresponding states $|\psi^{\perp}\rangle_1$. To verify this state preparation process, we map the prepared spinwave excitation out to a single photon (photon 2 in Fig. 1) by applying a read pulse on ensemble A and analyze its polarization using the quantum state tomography \cite{James2001}. The reconstructed six spinwave states ($\rho_i$ with $i = 1$ to $6$) of ensemble A  are plotted in the Bloch sphere as shown in Fig. 2. The average fidelity between the measured and ideal states is $97.5(2) \%$.

\begin{figure}[hbtp]
\centering
\includegraphics[width=0.5\textwidth]{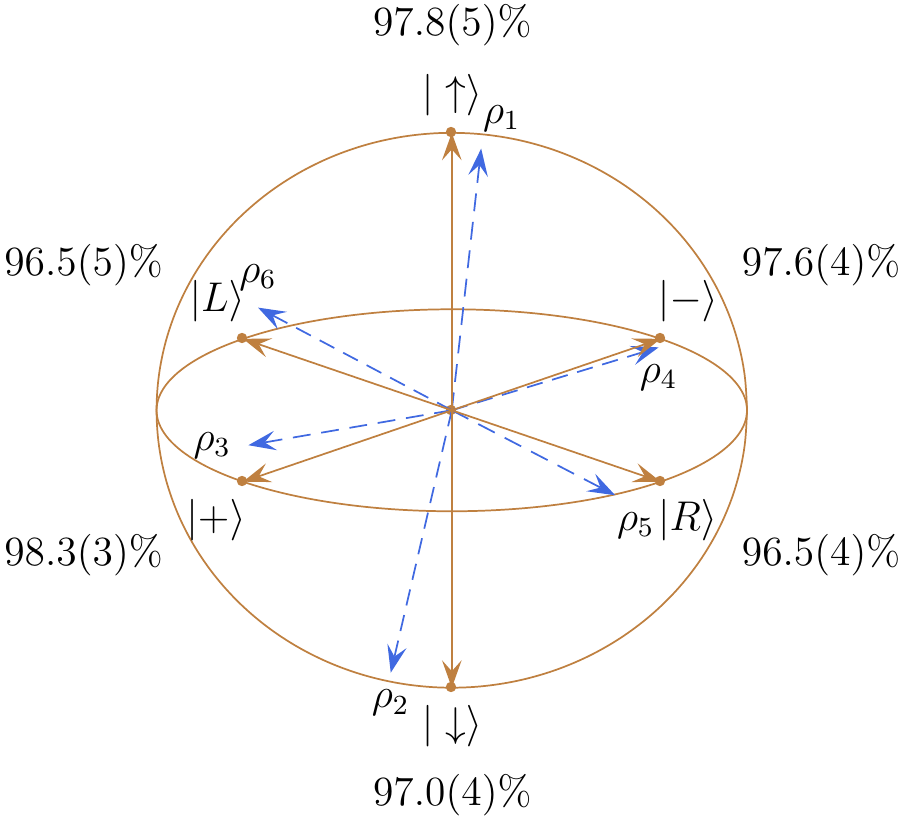}
\caption{\textbf{Bloch sphere representation of the tomography result for the prepared atomic states}. The solid arrow lines represent six target states ($|\uparrow\rangle$, $|\downarrow\rangle$, $|+\rangle$, $|-\rangle$, $|R\rangle$ and $|L\rangle$). The dashed arrow lines correspond to the six measured states ($\rho_i$ with $i = 1$ to $6$). Calculated fidelities between the measured and target states are shown, showing a near-perfect agreement between the two states. Errors for the fidelities are calculated based on the Poisson statistics of raw photon counts.}\label{tele:fig:preparation}
\end{figure}

Next, we establish the necessary quantum channel connecting the two atomic ensembles. The channel is in the form of entanglement between the spinwave state of atomic ensemble B (stationary and storable) and the polarization of a single photon which can be distributed far apart. In our experiment, it is created through the process of Raman scattering. Each time when a write pulse is applied, with a small probability, a pair of entangled state between the scattered photon 3 and the spinwave of ensemble B is generated in the form
\begin{equation}
|\Phi^+\rangle_{3B} = 1/\sqrt{2}(|H\rangle_3|\uparrow\rangle_B + |V\rangle_3|\downarrow\rangle_B).
\end{equation}
To test the robustness of our teleportation protocol over long distance, we send photon 3 to node A through a 150-m long single-mode fibre which has an intrinsic loss of about $11.4\,\%$. The temperature-dependent slow drift of polarization rotation caused by this fibre is actively checked and compensated.

To teleport the state $|\psi\rangle_{A}$ from node A to B, we need to make a joint Bell-state measurement (BSM) between  $|\psi\rangle_{A}$ and photon 3. It is, however, difficult to perform a direct BSM between a single photon and a spinwave. To remedy this problem, we convert the spinwave excitation in atomic ensemble A to a single photon (photon 2) by shining a strong read pulse. Before the conversion, in order to compensate the time delay of entanglement preparation in node B and transmission of photon 3 from node B to node A, the prepared state $|\psi\rangle_{A}$ is stored for 1.6 $\mu$s. The photons 2 and 3 are then superposed on a polarizing beam splitter (PBS) for BSM (see the setup in Fig. 1). Stable synchronization of these two independent narrow-band single photons which have coherence length of $\sim$$\,7.5$ meters is much easier compared to previous photonic teleportation experiments with parametric down-conversion where the coherence length of the photons is a few hundred micrometers \cite{Bouwmeester1997}, and thus extendable to a large-scale implementation. In addition to ensuring a good spatial and temporal overlap between the photon 2 and 3, their frequency should also be made indistinguishable. Thus, the $|g\rangle$ and $|s\rangle$ in the $\Lambda$ level schemes are arranged to be opposite between A and B, as shown in Fig. 1 as insets. The initial state where the atoms stay is also opposite. By coincidence detection and analysis of the two output photon polarization in the $|\pm\rangle$ basis \cite{pan1998}, we are able to discriminate two of them, \textit{i.e.}, $|\Phi^+\rangle_{23}$ and $|\Phi^-\rangle_{23}$. The classical measurement results are sent to node B. When we detect $|\Phi^+\rangle_{23}$, the teleportation is successful without further operation, while in case of $|\Phi^-\rangle_{23}$, a $\pi$ phase shift operation on $|\downarrow\rangle_B$ is required.

In order to evaluate the performance of the teleportation process, the teleported state in atomic ensemble B is measured by applying a read laser converting the spin-wave excitation to a single photon (photon 4 in Fig. 1) whose polarization is analyzed. Quantum state tomography for the teleported state is applied for all the six input states shown in Fig. 2. For the events in which BSM result is $|\Phi^-\rangle$, an artificial $\pi$ phase shift operation is applied to the reconstructed states. Based on this result, we calculate the fidelities between the prepared input states and the teleported states. Since in this case both the input and teleported states are mixed states in practice, we adopt the formula \cite{Gilchrist2005} of
$F(\rho_{1},\,\rho_{2}) \equiv \{\mathrm{tr}[(\sqrt{\rho_{1}}\rho_{2}\sqrt{\rho_{1}})^{1/2}]\}^2$ where $\rho_{1}$ and $\rho_{2}$ are arbitrary density matrices. Calculated results are listed in Tab. \ref{tele:tab:fiber}. We obtain an average fidelity of $F_{\mathrm{avg}}=95 \pm 1 \%$ which is well above the threshold of 2/3 attainable with classical means \cite{Massar1995}. Further, the state tomography results allow us to characterize the teleportation process using the technique of quantum process tomography \cite{O'Brien2004}. An arbitrary single-qubit operation on an input state $\rho_{\mathrm{in}}$ can be described by a process matrix $\chi$, which is defined as $\rho=\sum_{i,j=0}^{3}\chi_{ij}\hat{\sigma_i} \rho_{\mathrm{in}}\hat{\sigma_j}$ where $\rho$ is the output state and $\hat{\sigma_i}$ are Pauli matrices with $\hat{\sigma_0} = I$, $\hat{\sigma_1} = \hat{\sigma_x}$, $\hat{\sigma_2} = \hat{\sigma_y}$ and $\hat{\sigma_3} = \hat{\sigma_z}$. We use the maximum likelihood method \cite{O'Brien2004} to determine the most likely physical process matrix of our teleportation process. The measured process matrix is shown in Fig. 3. For an ideal teleportation process, there is only one nonzero element of $\chi_{00}^{\mathrm{ideal}}=1$. Therefore we get the calculated process fidelity of $F_{\mathrm{proc}}\equiv\mathrm{tr}(\chi \, \chi^{\mathrm{ideal}}) = 87(2)\%$ with the error calculated based on the Poisson distribution of original counts. The deviation from unit fidelity is mainly caused by the non-perfect entanglement of $|\Phi^+\rangle_{3B}$ and non-perfect interference on the PBS in the BSM stage.

\begin{table}
\centering \caption{Calculated fidelities between the prepared states of atomic ensemble A and the teleported states in atomic ensemble B based on the quantum state tomography results using the maximum likelihood method.}
\label{tele:tab:fiber}
\begin{tabular*}{1\columnwidth}{@{\extracolsep{\fill}}cccc}
\hline\hline
Input state of ensemble A & &Fidelity &\\
\hline
$\rho_1$ & &97(1)\% &\\
$\rho_2$ & &93(2)\% &\\
$\rho_3$ & &96(2)\% &\\
$\rho_4$ & &94(3)\% &\\
$\rho_5$ & &97(4)\% &\\
$\rho_6$ & &96(2)\% &\\
\hline
\end{tabular*}
\end{table}	

\begin{figure}[hbtp]
\centering
\includegraphics[width=0.49\columnwidth]{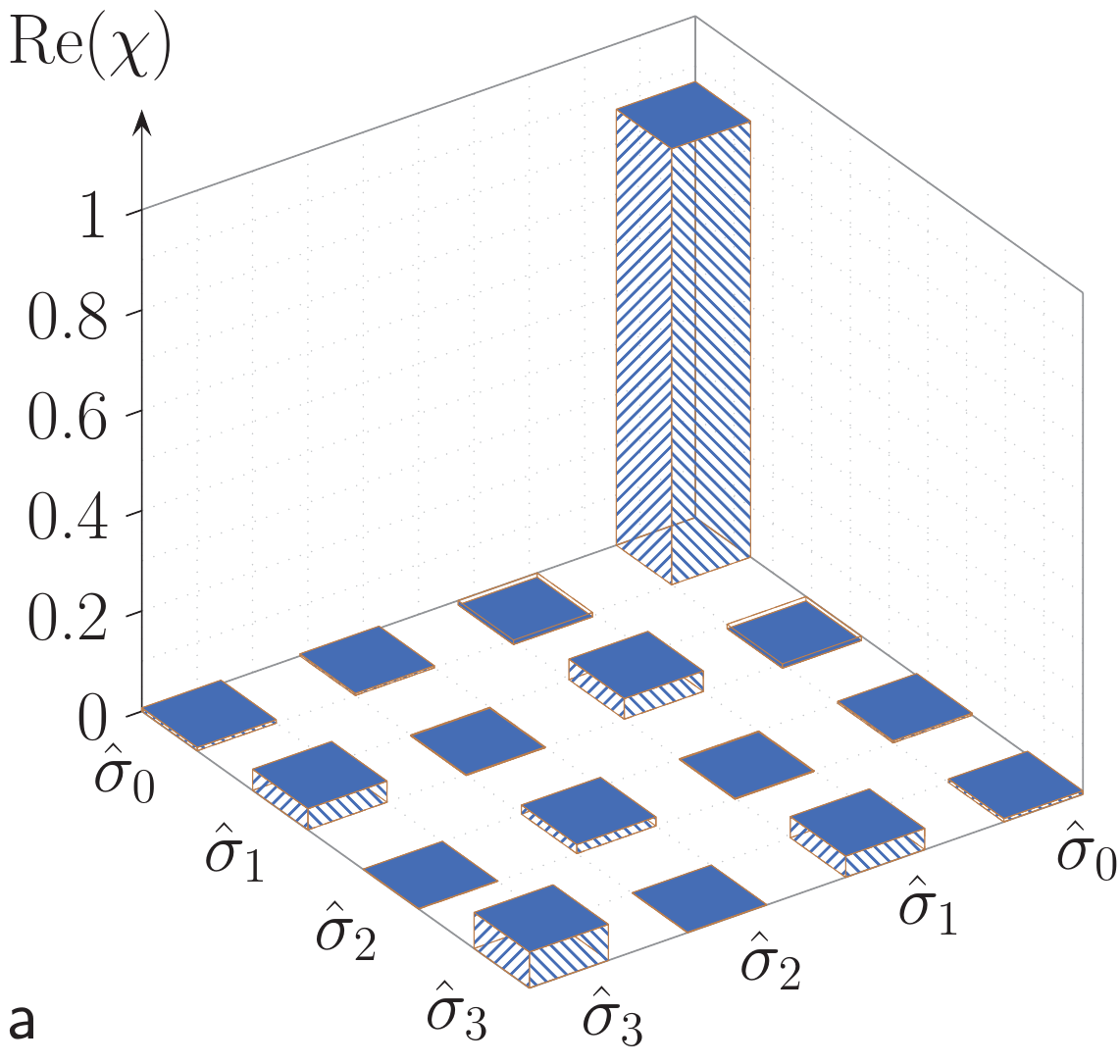}
\includegraphics[width=0.49\columnwidth]{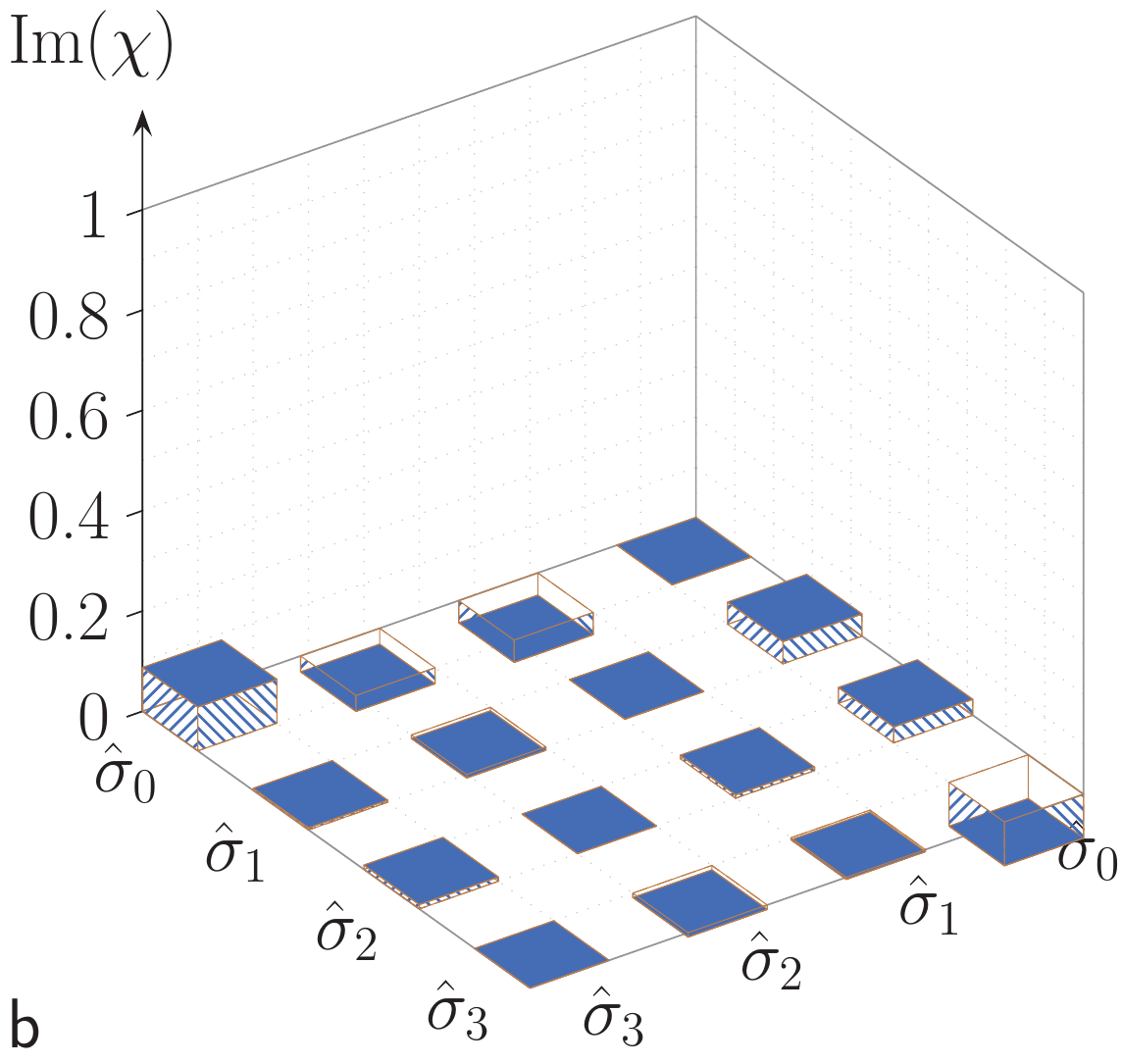}
\caption{\textbf{Measured process matrix $\chi$ for the teleportation}. The real part is shown in \textbf{a} and the imaginary part is shown in \textbf{b}. For an ideal teleportation process there should be only one nonzero element ($\chi_{00}=1$).}\label{tele:fig:process_tomo}
\end{figure}

We note that in our experiment the auxiliary entanglement pair between atomic ensemble B and photon 3 is probabilistic, thus our teleportation process also works probabilistically. For each input state, our teleportation process succeeds with a probability of $\eta_A P_B /2 \simeq 10^{-4}$, where $\eta_A$ (7\%) is the detected retrieval efficiency of ensemble A, $P_B$ ($3 \times 10^{-3}$) is the detection probability of a write-out photon from ensemble B during each write trial, and the $1/2$ is due to the efficiency of BSM (two Bell states out of four). The success probability is 4 orders of magnitude larger than the previous trapped-ion teleportation experiment \cite{Olmschenk2009}. Another useful feature of our experiment is that a trigger signal is available to herald the success of teleportation, which can benefit many applications including long-distance quantum communication \cite{Duan2001, gisin_rmp} and distributed quantum computing \cite{Kimble2008, qc_atomic_ensemble}. This trigger signal comes from the coincidence detection between $D_2$ and $D_3$ in the BSM stage. Let us further analyze the read-out noise of ensemble A and high order excitations of ensemble B. We find that the BSM signal is mixed with some noise which could give a fake trigger for teleportation. There are mainly three contributions for the BSM signal as listed below:
\begin{equation}
\begin{array}{lccc}
\rm{from} & \rm{A \& B} & \rm{A} & \rm{B} \\
\hline
\rm{probability~~} & \eta_A P_B & \eta_A P_A & {P_B}^2
\end{array}
\label{tele:eq:bsm}
\end{equation}
where $P_A$ is the detection probabilities of a write-out photon from ensemble A during each write trial. The first term is the desired term which corresponds to the case that one photon is retrieved out from node A and the other is the write-out photon from node B. The second term means that both photons are from node A, with one being the retrieved photon and the other the read-out noise photon which has a similar probability as the excitation probability. The third term comes from the case that both photons are from node B caused by double excitations. In order to have a high heralding fidelity, the proportion of the first term should be as high as possible, i.e., the following requirement should be fulfilled:
\begin{equation}
P_A \ll P_B \ll \eta_A.
\label{tele:eq:npt:cond}
\end{equation}
In our experiment $P_B \ll \eta_A$ is satisfied ($3 \times 10^{-3} \ll 7\%$).
In order to fulfill the first half inequality of Eq. \ref{tele:eq:npt:cond}, we reduce the excitation probability of ensemble A to $P_A \simeq 0.30 \times 10^{-3}$. Under this condition, we remeasure the teleported states for the six inputs and obtain an average post-selected fidelity of $F_{\mathrm{avg}}=93(2)\%$. This fidelity is slightly lower than the high excitation case (Tab. \ref{tele:tab:fiber}) due to the relatively higher contribution of background noise which mainly includes leakage of control laser (write, read, filter cell pumping beam, etc), stray light and detector dark counts. The fidelity of heralded teleportation, defined as  $F_{\mathrm{her}} \equiv F \, \eta_{\mathrm{her}}$ with the heralding efficiency $\eta_{\mathrm{her}} \equiv p_\mathrm{234}/p_\mathrm{23}\eta_B$ in which $p_\mathrm{234}$ and $p_\mathrm{23}$ are the joint detection probabilities of corresponding detectors conditioned on a detection event on D1, is measured to be $88(7)\%$ averaged over the six different input and output states. The imperfection of this heralding fidelity is mainly limited by the high-order excitations and background excitations. High-order excitations can be inhibited by making use of the Rydberg blockade effect \cite{simon10, Zhao2010}. Background excitations can be suppressed by putting the atomic ensemble inside an optical cavity so that the emission of scattered photons is enhanced only in predefined directions \cite{Simon2007}. These methods can in principle boost the heralding efficiency without lowering the excitation probability of ensemble A.

In summary, we have experimentally demonstrated heralded, high-fidelity quantum teleportation between two atomic ensembles linked by a 150-m long optical fibre using narrow-band single photons as quantum messengers. From a fundamental point of view \cite{Zeilinger2003}, this is interesting as the first teleportation between two macroscopic-sized objects \cite{macroscopic} at a distance of macroscopic scale. From a practical perspective, the combined techniques demonstrated here, including the heralded state preparation with feedback control, coherent mapping between matter and light, and quantum state teleportation, may provide a useful toolkit for quantum information transfer among different nodes in a quantum network \cite{Kimble2008, luming_rmp, gisin_rmp}. Moreover, these techniques could also be useful in the scheme for measurement-based quantum computing with atomic ensemble \cite{qc_atomic_ensemble}, \textit{e.g.}, to construct and connect atomic cluster states. Compared with the previous implementation with trapped ions \cite{Olmschenk2009}, for each input state, our experiment features a much higher (4 orders of magnitude) success probability. This is an advantage of the atomic ensembles where the collective enhancement enables efficient conversion of atomic qubits to photons in specific modes, avoiding the low efficiencies associated with the free space emission into the full solid angle in case of single ions \cite{Olmschenk2009}. Methods for further increasing the success probability include using a low-finesse optical cavity to improve the spinwave-to-photon conversion efficiency \cite{Simon2007} (higher $\eta_A$), and using the measurement-based scheme and another assisted ensemble to create the auxiliary photon-spinwave entanglement near deterministically (higher $P_B$) \cite{Chou2007,Yuan2008}. In the present experiment, the storage lifetime ($\sim$129 $\mu$s) of the prepared spinwave states in the quantum memories slightly exceeds the average time required ($\sim$97.5 $\mu$s) to create a pair of assistant remote entanglement for teleportation. The storage lifetime in the atomic ensembles can be increased up to 100 ms by making use of optical lattices to confine atomic motion \cite{Radnaev2010}. With these improvements we could envision quantum teleportation experiments among multiple atomic-ensemble nodes in the future.

\subsection*{Methods}
\textbf{Experimental details.}
Our experiment is operated with a repetition rate of 71.4 Hz. Within each cycle, the starting 11 ms is used to capture the atoms and cool them to $\sim$100 $\mu$K. The following 3 ms duration is used for the teleportation experiment, during which the trapping beams and the magnetic quadrupole field are switched off. Optical pumping to the Zeeman sublevel of $m_F=0$ is applied for ensemble A to increase the storage lifetime. Each write trial for ensemble A and B lasts for 3.38 $\mu$s and 975 ns, respectively. The probability to create a pair of photon-spinwave entanglement in node B within each write trial is about 0.01, thus the average time required to create a pair of assistant entanglement for teleportation is about 97.5 us. The write/read control pulses have a time duration of 50 ns, a beam waist of $\sim$240 $\mu$m. The write/read beams for both ensembles are on resonance with the corresponding transitions shown in Fig. 1. The polarization for the write (read) beams is vertical (horizontal), that is, perpendicular (in parallel) to the drawing plane in Fig. 1. The Rabi frequency for the write and read beams is 1.7 MHz and 14.6 MHz, respectively. The detection beam waist for the write-out  and read-out single-photons is $\sim$100 $\mu$m. The intersection angle between the write beam and the write-out photon mode for ensemble A(B) is $0.5^\circ$($3^\circ$). All the control pulse sequences are generated from a FPGA logic box. The output from single-photon detectors ($D_1$ to $D_4$) are either registered with a multi-channel time analyzer during the setup optimization, or with the logic box during data measurement for the teleportation process.

\textbf{Acknowledge}: This work was supported by the National Natural Science Foundation of China, the National Fundamental Research Program of China (Grant No. 2011CB921300), the Chinese Academy of Sciences, the Youth Qianren Program, the European Commission through the ERC Grant, and the STREP project HIP.


\begin{thebibliography}{10}

\bibitem{Gisin2002}
Gisin, N, Ribordy, G, Tittel, W,  \& Zbinden, H.
\newblock (2002) Quantum cryptography.
\newblock {\em Rev. Mod. Phys.} {\bf 74}, 145--195.

\bibitem{physicalreport}
Yuan, Z.-S, Bao, X.-H, Lu, C.-Y, Zhang, J, Peng, C.-Z,  \& Pan, J.-W.
\newblock (2010) Entangled photons and quantum communication.
\newblock {\em Phys. Rep.} {\bf 497}, 1--40.

\bibitem{Stucki2009}
Stucki, D, Walenta, N, Vannel, F, Thew, R.~T, Gisin, N, Zbinden, H, Gray, S,
  Towery, C.~R,  \& Ten, S.
\newblock (2009) High rate, long-distance quantum key distribution over 250 km
  of ultra low loss fibres.
\newblock {\em New J. Phys.} {\bf 11}, 075003.

\bibitem{Liu2010}
Liu, Y, Chen, T.-Y, Wang, J, Cai, W.-Q, Wan, X, Chen, L.-K, Wang, J.-H, Liu,
  S.-B, Liang, H, Yang, L, Peng, C.-Z, Chen, K, Chen, Z.-B,  \& Pan, J.-W.
\newblock (2010) Decoy-state quantum key distribution with polarized photons
  over 200 km.
\newblock {\em Opt. Express} {\bf 18}, 8587--8594.

\bibitem{Briegel1998}
Briegel, H.~J, Dur, W, Cirac, J.~I,  \& Zoller, P.
\newblock (1998) Quantum repeaters: The role of imperfect local operations in
  quantum communication.
\newblock {\em Phys. Rev. Lett.} {\bf 81}, 5932--5935.

\bibitem{Duan2001}
Duan, L.-M, Lukin, M.~D, Cirac, J.~I,  \& Zoller, P.
\newblock (2001) Long-distance quantum communication with atomic ensembles and
  linear optics.
\newblock {\em Nature} {\bf 414}, 413--418.

\bibitem{Kimble2008}
Kimble, H.~J.
\newblock (2008) The quantum internet.
\newblock {\em Nature} {\bf 453}, 1023--1030.

\bibitem{luming_rmp}
Duan, L.-M \& Monroe, C.
\newblock (2010) Colloquium: Quantum networks with trapped ions.
\newblock {\em Rev. Mod. Phys.} {\bf 82}, 1209--1224.

\bibitem{gisin_rmp}
Sangouard, N, Simon, C, de~Riedmatten, H,  \& Gisin, N.
\newblock (2011) Quantum repeaters based on atomic ensembles and linear optics.
\newblock {\em Rev. Mod. Phys.} {\bf 83}, 33--80.

\bibitem{simon2010}
Simon, C, Afzelius, M, Appel, J, Boyer de~la Giroday, A, Dewhurst, S.~J, Gisin,
  N, Hu, C.~Y, Jelezko, F, Kr\"{o}ll, S, M\"{u}ller, J.~H, Nunn, J, Polzik,
  E.~S, Rarity, J.~G, De~Riedmatten, H, Rosenfeld, W, Shields, A.~J, Sk\"{o}ld,
  N, Stevenson, R.~M, Thew, R, Walmsley, I.~A, Weber, M.~C, Weinfurter, H,
  Wrachtrup, J,  \& Young, R.~J.
\newblock (2010) Quantum memories.
\newblock {\em Eur. Phys. J. D} {\bf 58}, 1--22.

\bibitem{teleportation_Bennett}
Bennett, C.~H, Brassard, G, Cr\'epeau, C, Jozsa, R, Peres, A,  \& Wootters,
  W.~K.
\newblock (1993) Teleporting an unknown quantum state via dual classical and
  $\mathrm{E}$instein-$\mathrm{P}$odolsky-$\mathrm{R}$osen channels.
\newblock {\em Phys. Rev. Lett.} {\bf 70}, 1895--1899.

\bibitem{kuzmich}
Kuzmich, A, Bowen, W.~P, Boozer, A.~D, Boca, A, Chou, C.~W, Duan, L.-M,  \&
  Kimble, H.~J.
\newblock (2003) Generation of nonclassical photon pairs for scalable quantum
  communication with atomic ensembles.
\newblock {\em Nature} {\bf 423}, 731--734.

\bibitem{lukin}
van~der Wal, C.~H, Eisaman, M.~D, Andre, A, Walsworth, R.~L, Phillips, D.~F,
  Zibrov, A.~S,  \& Lukin, M.~D.
\newblock (2003) Atomic memory for correlated photon states.
\newblock {\em Science} {\bf 301}, 196--200.

\bibitem{Choi2008}
Choi, K.~S, Deng, H, Laurat, J,  \& Kimble, H.~J.
\newblock (2008) Mapping photonic entanglement into and out of a quantum
  memory.
\newblock {\em Nature} {\bf 452}, 67--71.

\bibitem{Chou2007}
Chou, C.-W, Laurat, J, Deng, H, Choi, K.~S, de~Riedmatten, H, Felinto, D,  \&
  Kimble, H.~J.
\newblock (2007) Functional quantum nodes for entanglement distribution over
  scalable quantum networks.
\newblock {\em Science} {\bf 316}, 1316--1320.

\bibitem{Yuan2008}
Yuan, Z.-S, Chen, Y.-A, Zhao, B, Chen, S, Schmiedmayer, J,  \& Pan, J.-W.
\newblock (2008) Experimental demonstration of a $\mathrm{BDCZ}$ quantum
  repeater node.
\newblock {\em Nature} {\bf 454}, 1098--1101.

\bibitem{Radnaev2010}
Radnaev, A.~G, Dudin, Y.~O, Zhao, R, Jen, H.~H, Jenkins, S.~D, Kuzmich, A,  \&
  Kennedy, T. A.~B.
\newblock (2010) A quantum memory with telecom-wavelength conversion.
\newblock {\em Nature Phys.} {\bf 6}, 894--899.

\bibitem{macroscopic}
Julsgaard, B, Kozhekin, A,  \& Polzik, E.~S.
\newblock (2001) Experimental long-lived entanglement of two macroscopic
  objects.
\newblock {\em Nature} {\bf 413}, 400--403.

\bibitem{Chou2005}
Chou, C.~W, de~Riedmatten, H, Felinto, D, Polyakov, S.~V, van Enk, S.~J,  \&
  Kimble, H.~J.
\newblock (2005) Measurement-induced entanglement for excitation stored in
  remote atomic ensembles.
\newblock {\em Nature} {\bf 438}, 828--832.

\bibitem{Bouwmeester1997}
Bouwmeester, D, Pan, J.-W, Mattle, K, Eibl, M, Weinfurter, H,  \& Zeilinger, A.
\newblock (1997) Experimental quantum teleportation.
\newblock {\em Nature} {\bf 390}, 575--579.

\bibitem{Martini1997}
Boschi, D, Branca, S, De~Martini, F, Hardy, L,  \& Popescu, S.
\newblock (1998) Experimental realization of teleporting an unknown pure
  quantum state via dual classical and
  $\mathrm{E}$instein-$\mathrm{P}$odolsky-$\mathrm{R}$osen channels.
\newblock {\em Phys. Rev. Lett.} {\bf 80}, 1121--1125.

\bibitem{TeleCWFurusawa1998}
Furusawa, A, S$\phi$rensen, J.~L, Braunstein, S.~L, Fuchs, C.~A, Kimble, H.~J,
  \& Polzik, E.~S.
\newblock (1998) Unconditional quantum teleportation.
\newblock {\em Science} {\bf 282}, 706--709.

\bibitem{Sherson2006}
Sherson, J.~F, Krauter, H, Olsson, R.~K, Julsgaard, B, Hammerer, K, Cirac, I,
  \& Polzik, E.~S.
\newblock (2006) Quantum teleportation between light and matter.
\newblock {\em Nature} {\bf 443}, 557--560.

\bibitem{Chen2008}
Chen, Y.-A, Chen, S, Yuan, Z.-S, Zhao, B, Chuu, C.-S, Schmiedmayer, J,  \& Pan,
  J.-W.
\newblock (2008) Memory-built-in quantum teleportation with photonic and atomic
  qubits.
\newblock {\em Nature Phys.} {\bf 4}, 103--107.

\bibitem{Riebe2004}
Riebe, M, Haffner, H, Roos, C.~F, Hansel, W, Benhelm, J, Lancaster, G. P.~T,
  Korber, T.~W, Becher, C, Schmidt-Kaler, F, James, D. F.~V,  \& Blatt, R.
\newblock (2004) Deterministic quantum teleportation with atoms.
\newblock {\em Nature} {\bf 429}, 734--737.

\bibitem{Barrett2004}
Barrett, M.~D, Chiaverini, J, Schaetz, T, Britton, J, Itano, W.~M, Jost, J.~D,
  Knill, E, Langer, C, Leibfried, D, Ozeri, R,  \& Wineland, D.~J.
\newblock (2004) Deterministic quantum teleportation of atomic qubits.
\newblock {\em Nature} {\bf 429}, 737--739.

\bibitem{Olmschenk2009}
Olmschenk, S, Matsukevich, D.~N, Maunz, P, Hayes, D, Duan, L.-M,  \& Monroe, C.
\newblock (2009) Quantum teleportation between distant matter qubits.
\newblock {\em Science} {\bf 323}, 486--489.

\bibitem{Fleischhauer2005}
Fleischhauer, M, Imamoglu, A,  \& Marangos, J.~P.
\newblock (2005) Electromagnetically induced transparency: Optics in coherent
  media.
\newblock {\em Rev. Mod. Phys.} {\bf 77}, 633.

\bibitem{Simon2007}
Simon, J, Tanji, H, Thompson, J.~K,  \& Vuleti\ifmmode~\acute{c}\else
  \'{c}\fi{}, V.
\newblock (2007) Interfacing collective atomic excitations and single photons.
\newblock {\em Phys. Rev. Lett.} {\bf 98}, 183601.

\bibitem{Bao2012}
Bao, X.-H, Reingruber, A, Dietrich, P, Rui, J, Duck, A, Strassel, T, Li, L,
  Liu, N.~L, Zhao, B,  \& Pan, J.-W.
\newblock (2012) Efficient and long-lived quantum memory with cold atoms inside
  a ring cavity.
\newblock {\em Nature Phys.} {\bf 8}, 517--521.

\bibitem{RSP_Bennett}
Bennett, C.~H, DiVincenzo, D.~P, Shor, P.~W, Smolin, J.~A, Terhal, B.~M,  \&
  Wootters, W.~K.
\newblock (2001) Remote state preparation.
\newblock {\em Phys. Rev. Lett.} {\bf 87}, 077902.

\bibitem{Chen2007}
Chen, S, Chen, Y.-A, Zhao, B, Yuan, Z.-S, Schmiedmayer, J,  \& Pan, J.-W.
\newblock (2007) Demonstration of a stable atom-photon entanglement source for
  quantum repeaters.
\newblock {\em Phys. Rev. Lett.} {\bf 99}, 180505.

\bibitem{Zhao2009}
Zhao, B, Chen, Y.-A, Bao, X.-H, Strassel, T, Chuu, C.-S, Jin, X.-M,
  Schmiedmayer, J, Yuan, Z.-S, Chen, S,  \& Pan, J.-W.
\newblock (2009) A millisecond quantum memory for scalable quantum networks.
\newblock {\em Nature Phys.} {\bf 5}, 95--99.

\bibitem{James2001}
James, D. F.~V, Kwiat, P.~G, Munro, W.~J,  \& White, A.~G.
\newblock (2001) Measurement of qubits.
\newblock {\em Phys. Rev. A} {\bf 64}, 052312.

\bibitem{pan1998}
Pan, J.-W \& Zeilinger, A.
\newblock (1998) Greenberger-$\mathrm{H}$orne-$\mathrm{Z}$eilinger-state
  analyzer.
\newblock {\em Phys. Rev. A} {\bf 57}, 2208.

\bibitem{Gilchrist2005}
Gilchrist, A, Langford, N.~K,  \& Nielsen, M.~A.
\newblock (2005) Distance measures to compare real and ideal quantum processes.
\newblock {\em Phys. Rev. A} {\bf 71}, 062310.

\bibitem{Massar1995}
Massar, S \& Popescu, S.
\newblock (1995) Optimal extraction of information from finite quantum
  ensembles.
\newblock {\em Phys. Rev. Lett.} {\bf 74}, 1259--1263.

\bibitem{O'Brien2004}
O'Brien, J.~L, Pryde, G.~J, Gilchrist, A, James, D. F.~V, Langford, N.~K,
  Ralph, T.~C,  \& White, A.~G.
\newblock (2004) Quantum process tomography of a controlled-$\mathrm{NOT}$
  gate.
\newblock {\em Phys. Rev. Lett.} {\bf 93}, 080502.

\bibitem{qc_atomic_ensemble}
Barrett, S.~D, Rohde, P.~P,  \& Stace, T.~M.
\newblock (2010) Scalable quantum computing with atomic ensembles.
\newblock {\em New J. Phys.} {\bf 12}, 093032.

\bibitem{simon10}
Han, Y, He, B, Heshami, K, Li, C.-Z,  \& Simon, C.
\newblock (2010) Quantum repeaters based on $\mathrm{R}$ydberg-blockade-coupled
  atomic ensembles.
\newblock {\em Phys. Rev. A} {\bf 81}, 052311.

\bibitem{Zhao2010}
Zhao, B, Muller, M, Hammerer, K,  \& Zoller, P.
\newblock (2010) Efficient quantum repeater based on deterministic
  $\mathrm{R}$ydberg gates.
\newblock {\em Phys. Rev. A} {\bf 81}, 052329.

\bibitem{Zeilinger2003}
Zeilinger, A.
\newblock (2003) Quantum teleportation.
\newblock {\em Sci. Am.} {\bf 13}, 34--43.

\end{thebibliography}

\end{document}